\documentclass{ws-p8-50x6-00}

\newcommand{\hbo}{\hbar \omega}  
          
\newcommand{\lm}{(\lambda,\mu)}
    
\newcommand{\lms}{(\lambda_{\sigma},\mu_{\sigma})}

\begin{document}

\title{Partial Dynamical Symmetry in a Fermionic Many-Body System 
}

\author{Jutta Escher}

\address{TRIUMF, 4004 Wesbrook Mall, Vancouver, B.C. V6T 2A3, Canada\\
E-mail: escher@triumf.ca}


\maketitle

\abstracts{
The concept of partial symmetry is introduced for an interacting fermion
system.  The associated Hamiltonians are shown to be closely related to
a realistic nuclear quadrupole-quadrupole interaction. An application
to $^{12}$C is presented.
}

\section{Introduction}

The fundamental concept underlying algebraic theories in quantum physics is
that of an exact or dynamical symmetry.
Realistic quantum systems, however, often require the associated symmetry
to be broken in order to allow for a proper description of some observed
basic features.
Partial dynamical symmetry (PDS) describes an intermediate situation in which
some eigenstates exhibit a symmetry which the associated Hamiltonian
does not share.
The objective of this approach is to remove undesired constraints from the
theory while preserving the useful aspects of a dynamical symmetry, such
as solvability, for a subset of eigenstates.\cite{GenPDS,Escher00a}
Here we present an example of a PDS in an interacting fermion system.
In the symplectic shell model of nuclei,\cite{SymplM} we introduce PDS 
Hamiltonians which are closely related to the nuclear quadrupole-quadrupole 
interaction.  An application to $^{12}$C is discussed. 

\section{PDS Hamiltonians and quadrupole-quadrupole interaction}

The quadrupole-quadrupole interaction is an important ingredient in models
that aim at reproducing quadrupole collective properties of nuclei.
A model which is able to fully accommodate the action of the collective
quadrupole operator,
$Q_{2m}=\sqrt{16\pi/5} \sum_s r^2_s Y_{2m} (\hat{r}_s)$,
is the symplectic shell model (SSM), an algebraic scheme which respects the
Pauli principle.\cite{SymplM}
In the SSM, this operator takes the form
$Q_{2m} = \sqrt{3} ( \hat{C}^{(11)}_{2m}
+ \hat{A}^{(20)}_{2m} + \hat{B}^{(02)}_{2m} )$,
where $\hat{A}^{(20)}_{lm},\,\hat{B}^{(02)}_{lm}$,
and $\hat{C}^{(11)}_{lm}$ are symplectic generators with good
SU(3) [superscript $(\lambda,\mu)$]
and SO(3) [subscript $l,m$] tensorial properties.
The $\hat{A}^{(20)}_{lm}$ ($\hat{B}^{(02)}_{lm}$), $l$ = 0 or 2,
create (annihilate) $2 \hbo$ excitations in the system.
The $\hat{C}^{(11)}_{lm}$, $l$ = 1 or 2, generate a SU(3) subgroup and
act only {\em within} one harmonic oscillator (h.o.) shell
($\sqrt{3} \hat{C}^{(11)}_{2m}=$ $Q^E_{2m}$, the quadrupole
operator of Elliott, which does not couple different
h.o.\ shells,\cite{Elliott58} and $\hat{C}^{(11)}_{1m}=\hat{L}_m$,
the angular momentum operator).        
The symplectic basis is generated by repeated application of $\hat{A}^{(20)}$
to a $0\hbo$ shell model configuration, labeled by its Elliott quantum
numbers $\lms$.
The resulting $N\hbo$ excited states ($N$=0,2,$\ldots$) are coupled to good
SU(3)$\supset$SO(3) symmetry $\lm \kappa L M$, where $\kappa$ enumerates
multiple occurrences of $L$ in the SU(3) irrep $\lm$.
This labeling scheme defines a dynamical symmetry basis.

The quadrupole-quadrupole interaction connects h.o.\ states differing
in energy by $0\hbo$, $\pm 2\hbo$, and $\pm 4\hbo$, and may be written as
\begin{eqnarray} 
Q_2 \cdot Q_2 &=& 9 \hat{C}_{SU3} - 3 \hat{C}_{Sp6} +
\hat{H}_0^2 - 2 \hat{H}_0 - 3 \hat{L}^2
- 6 \hat{A}_0 \hat{B}_0 \nonumber \\
&& + \{ \mbox{terms coupling different h.o.\/ shells} \} \; , \label{Eq:QQ}
\end{eqnarray}
where  $\hat{C}_{SU3}$ and $\hat{C}_{Sp6}$ are Casimir invariants of SU(3) 
and Sp(6,R). 
These operators, as well as $\hat{H}_0$ and $\hat{L}^2$,
are diagonal in the dynamical symmetry basis.
Unlike the Elliott quadrupole-quadrupole interaction, $Q_2 \cdot Q_2$ 
breaks SU(3) symmetry within each h.o.\ shell since 
$\hat{A}_0 \hat{B}_0$ mixes different SU(3) irreps.
To study the action of $Q_2 \cdot Q_2$ within such a shell, we
consider the Hamiltonians
\begin{eqnarray} 
\lefteqn{H(\beta_0,\beta_2) = \beta_0 \hat{A}_0 \hat{B}_0
+ \beta_2 \hat{A}_2 \cdot \hat{B}_2 }
\label{Eq:Hpds} \\
&& = \frac{\beta_2}{18} ( 9\hat{C}_{SU3} - 9\hat{C}_{Sp6}
+ 3\hat{H}_0^2 - 36\hat{H}_0 )
   + ( \beta_0 - \beta_2) \hat{A}_0 \hat{B}_0 \; .  \nonumber
\end{eqnarray}
For $\beta_0=\beta_2$, one recovers the dynamical symmetry, and for 
$\beta_0=12$, $\beta_2=18$, one obtains
$Q_2 \cdot Q_2 = H(\beta_0=12,\beta_2=18)+ const(N) - 3 \hat{L}^2$
+ terms coupling different shells, where $const(N)$ is constant for a
given h.o. $N\hbo$ excitation.  

For general $\beta_0 \neq \beta_2$, $H(\beta_0,\beta_2)$ exhibits partial 
SU(3) symmetry:
The Hamiltonian is not SU(3) invariant, yet it possesses a subset of
`special' states which respect the symmetry:                    
All 0$\hbo$ states are unmixed and span the entire $\lms$ irrep. 
Moreover, among the excited configurations ($N > 0$), one finds additional
states with good SU(3) symmetry. 
Unlike the $0\hbo$ states, however, they span only part of the corresponding 
SU(3) irreps.
There are other states at each excited level which do not preserve the
symmetry and therefore contain a mixture of irreps.
The partial SU(3) symmetry of $H(\beta_0,\beta_2)$ is converted into partial
dynamical symmetry by adding to it SO(3) rotation terms which lead
to $L(L+1)$-type splitting but do not affect the wave functions.
The solvable states then form rotational bands and since their wave
functions are known, one can evaluate their energies and the E2 rates between 
them analytically.\cite{Escher00b}

\section{Application to $^{12}$C}

\begin{figure}[t]
\hspace{1cm}
\epsfxsize=20pc 
\epsfbox{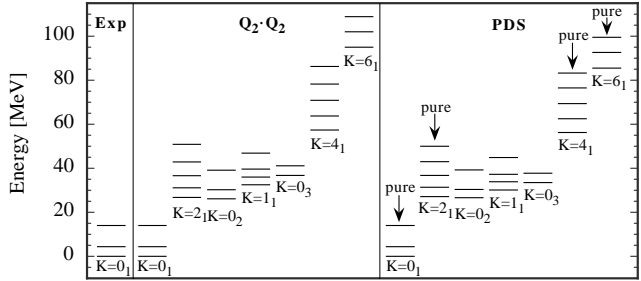} 
\caption{Energy spectra for $^{12}$C.
K=$0_1$ indicates the ground band in all three parts of the figure.
In addition, resonance bands dominated by 2$\hbo$ (K=$2_1,0_2,1_1,0_3$), 4$\hbo$ 
(K=$4_1$), and 6$\hbo$ excitations (K=$6_1$) are shown for the two calculations.
The angular momenta of the states in the rotational bands are $L$=0,2,4,$\ldots$ for
K=0 and $L$=K,K+1,K+2, $\ldots$ otherwise.
\label{Energies_C12}}
\vspace{-0.5cm} 
\end{figure}

To illustrate that the PDS Hamiltonians introduced here are physically
relevant, we present an application to $^{12}$C.
In Fig.~\ref{Energies_C12}, we compare the energy spectra of $H_{PDS} = h(N) + 
\xi H(\beta_{0}=12,\beta_{2}=18) + \gamma_2 \hat{L}^2 + \gamma_4 \hat{L}^4$
and $H_{Q \cdot Q} = \hat{H}_0 - \chi Q_2 \cdot Q_2 + d_2 \hat{L}^2 + d_4 \hat{L}^4$,
where $h(N)$ is constant for a given $N\hbo$ excitation.      
$H_{PDS}$ has families of pure SU(3) eigenstates which can be organized into 
rotational bands; they are indicated in the figure.

\begin{figure}[t]
\epsfxsize=28pc 
\epsfbox{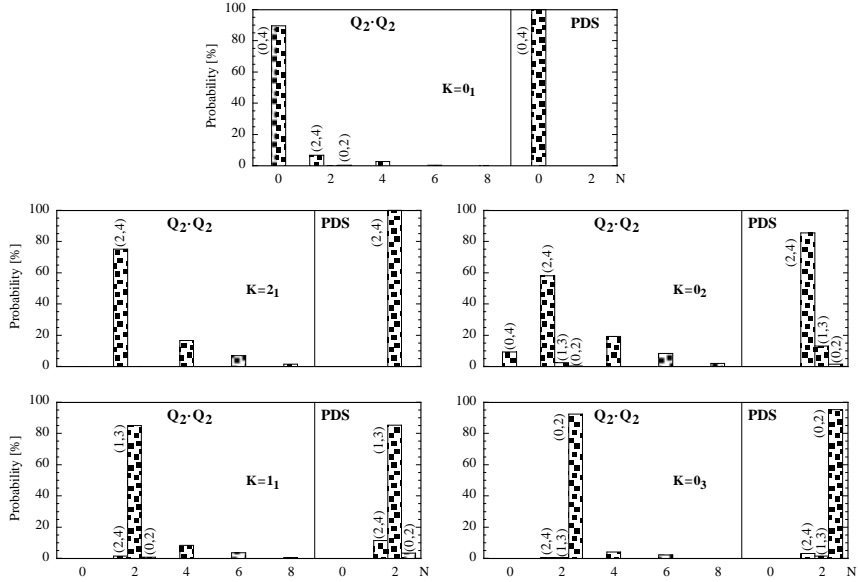} 
\vspace{-8.0cm}                                                                         
\caption{
Decompositions for calculated $L^{\pi}=2^+$ states of $^{12}$C.
Individual contributions from the relevant SU(3) irreps at the 0$\hbo$
and 2$\hbo$ levels are shown for both the $8\hbo$ $Q_2 \cdot Q_2$ calculation
and the PDS calculation.
In addition, the total strengths contributed by the $N\hbo$ excitations
for $N>2$ are given for the $Q_2 \cdot Q_2$ case.                          
\label{Decomp_C12_L2}}
\vspace{-0.5cm}
\end{figure}

Although the PDS Hamiltonian cannot account for intershell correlations,
it is able to reproduce various features of the quadrupole-quadrupole interaction,
as can be seen in Fig.~\ref{Decomp_C12_L2}, where the structure of selected PDS
eigenstates is compared to that of the corresponding $Q_2 \cdot Q_2$ eigenstates: 
PDS eigenfunctions do not contain admixtures from different $N\hbo$
configurations, but belong entirely to one level of excitation.
We find that, for reasonable interaction parameters, the $N\hbo$ level
to which a particular PDS band belongs is also dominant in the corresponding
band of exact $Q_2 \cdot Q_2$ eigenstates.
Moreover, within this dominant excitation, eigenstates of both Hamiltonians
have similar SU(3) distributions.
Structural differences, nevertheless, do arise and are reflected in the very
sensitive interband transition rates.\cite{Escher00b}
Overall, however, we may conclude that PDS eigenstates approximately
reproduce the structure of the exact $Q_2 \cdot Q_2$ eigenstates, for both
ground and the resonance bands.

\section{Summary}

The notion of partial dynamical symmetries extends the familiar concepts of 
exact and dynamical symmetries.
It is applicable when a subset of states exhibit a symmetry which
does not arise from invariance properties of the relevant Hamiltonian.
Recent studies, including the one presented here, show that partial
symmetries are indeed realized in various quantum systems.

\section*{Acknowledgments}
This research was carried out in collaboration with A.~Leviatan
(Hebrew U.).

\end{document}